\documentclass[a4paper]{spie}  

\usepackage{amsmath,amsfonts,amssymb,aas_macros}
\usepackage{graphicx}
\usepackage[colorlinks=true, allcolors=blue]{hyperref}

\title{The CUbesat Solar Polarimeter (CUSP): mission overview II}

\author[a]{Sergio~Fabiani}
\author[a]{Ettore~Del~Monte}
\author[a,b]{Andrea~Alimenti}
\author[g,n]{Riccardo~Campana}
\author[h]{Mauro~Centrone}
\author[a]{Enrico~Costa}
\author[a]{Nicolas~De~Angelis}
\author[g]{Giovanni~De~Cesare}
\author[a]{Sergio~Di~Cosimo}
\author[a]{Giuseppe~Di~Persio}
\author[a]{Abhay~Kumar}
\author[a]{Alessandro~Lacerenza}
\author[a]{Pasqualino~Loffredo}
\author[a,c]{Giovanni~Lombardi}
\author[a]{Lorenzo~Marra}
\author[l]{Gabriele~Minervini}
\author[a]{Fabio~Muleri}
\author[m]{Paolo~Romano}
\author[a]{Alda~Rubini}
\author[a]{Emanuele~Scalise}
\author[a,b]{Enrico~Silva}
\author[a]{Paolo~Soffitta}
\author[d]{Davide~Albanesi}
\author[e]{Ilaria~Baffo}
\author[f]{Daniele~Brienza}
\author[d]{Valerio~Campamaggiore}
\author[i]{Giovanni~Cucinella}
\author[j]{Andrea~Curatolo}
\author[d]{Giulia~de~Iulis}
\author[d]{Andrea~Del~Re}
\author[i]{Vito~Di~Bari}
\author[i]{Simone~Di~Filippo}
\author[f]{Immacolata~Donnarumma}
\author[e]{Pierluigi~Fanelli}
\author[k]{Nicolas~Gagliardi}
\author[d]{Paolo~Leonetti}
\author[f]{Matteo~Mergè}
\author[j,k]{Dario~Modenini}
\author[i]{Andrea~Negri}
\author[j]{Daniele~Pecorella}
\author[i]{Massimo~Perelli}
\author[k]{Alice~Ponti}
\author[d]{Francesca~Sbop}
\author[j,k]{Paolo~Tortora}
\author[f]{Alessandro~Turchi}
\author[f]{Valerio~Vagelli}
\author[f]{Emanuele~Zaccagnino}
\author[d]{Alessandro~Zambardi}
\author[e]{Costantino~Zazza}

\affil[a]{INAF-IAPS\\ via del Fosso del Cavaliere 100, 00133, Rome, Italy}
\affil[b]{Department of Industrial, Electronic and Mechanical Engineering, "Roma Tre" University, via V. Volterra 62, 00146, Rome, Italy}
\affil[c]{Department of Enterprise Engineering "Mario Lucenti”, University of Rome "Tor Vergata", Via Cracovia 50, 00133, Rome, Italy}
\affil[d]{DEDA Connect s.r.l.\\ via Vincenzo Lamaro 51, 00173 Rome, Italy}
\affil[e]{DEIM, University of "La Tuscia", Largo dell’Università, 01100, Viterbo, Italy}
\affil[f]{ASI, via del Politecnico snc\\ 00133, Rome, Italy}
\affil[g]{INAF-OAS Bologna\\ via Gobetti 93/3, 40129, Bologna, Italy}
\affil[h]{INAF-OAR\\ via Frascati 33, 00040, Monte Porzio Catone, Italy}
\affil[i]{IMT s.r.l.\\ via Carlo Bartolomeo Piazza 30, 00161, Rome, Italy}
\affil[j]{Department of Industrial Engineering - Alma Mater Studiorum Università di Bologna - Via Montaspro 97, 47121 Forlì, Italy}
\affil[k]{Interdepartmental Centre for Industrial Aerospace Research - Alma Mater Studiorum Università di Bologna -  Via Carnaccini 12, 47121 Forlì, Italy}
\affil[l]{INAF-Headquarters\\ viale del Parco Mellini 84, 00136, Rome, Italy}
\affil[m]{INAF-OACT\\ Via S. Sofia 78, 95123, Catania, Italy}
\affil[n]{INFN Sezione di Bologna, viale Berti Pichat 6/2,  40127, Bologna, Italia}

\authorinfo{Further author information: (Send correspondence to Sergio Fabiani)\\ E-mail: sergio.fabiani@inaf.it, Telephone: +39 064993 4450}

\begin{document} 
\maketitle

\begin{abstract}
The CUbesat Solar Polarimeter (CUSP) project is an Earth-orbiting CubeSat mission designed to measure the linear polarization of solar flares in the hard X-ray band using a Compton scattering polarimeter. CUSP will enable the study of magnetic reconnection and particle acceleration within the Sun's flaring magnetic structures. This project is being developed within the framework of the Italian Space Agency's Alcor Program\footnote{\url{https://www.asi.it/en/technologies-and-engineering/micro-and-nanosatellites/alcor-program/}, consulted
on 12 Jul 2025}, which aims to foster new CubeSat missions. CUSP entered its Phase B in December 2024, a phase scheduled to last 12 months. This paper reports on the current status of the CUSP mission design, mission analysis, and payload scientific performance.
\end{abstract}


\section{INTRODUCTION}
\label{sec:intro}

Solar flares (SFs) are powerful energetic phenomena occurring on the Sun that can significantly impact human activities both on Earth and in space. Solar activity, including SFs, can degrade radio communications, cause radio blackouts, and interfere with GNSS infrastructures and satellite communications. Furthermore, high-energy particles (protons and electrons) can deposit energy in satellite electronics, leading to malfunctions or even satellite loss. The occurrence of SFs is also linked to Coronal Mass Ejections (CMEs) and Solar Energetic Particle (SEP) events observed on Earth \cite{Papaioannou2016}. The CUbesat Solar Polarimeter (CUSP) aims to study SFs by measuring the linear polarization of their X-ray emission. The outcomes from CUSP are intended to enhance our understanding of the underlying physics during SFs and contribute to current and future Space Weather monitoring networks.

In the classical model of an SF, magnetic reconnection at the apex of a magnetic loop triggers a massive energy release. Particles are then accelerated along magnetic field lines towards the lower layers of the solar atmosphere and into interplanetary space. The energy spectrum of an SF below 100 keV is dominated by three main components:
\begin{itemize}
\item emission lines below 10 keV;
\item thermal Bremsstrahlung (expected to be weakly polarized) \cite{Emslie1980a};
\item non-thermal Bremsstrahlung, emerging from approximately 10–30 keV \cite{Nagasawa2022}. This component is primarily of interest for CUSP, because it is expected to be highly polarized \cite{Zharkova2010}.
\end{itemize}
During the early impulsive phase, the energy spectrum is characterized by significant non-thermal emission. Subsequent plasma heating leads to an increase in the thermal emission component, which typically surpasses the non-thermal emission in the 10-30 keV energy range, thereby diluting the overall polarization degree \cite{Nagasawa2022, Grigis2004, Dennis2005}.

Theoretical models predict high linear polarization in X-rays, depending on particle beaming and magnetic field properties \cite{Zharkova2010, Jeffrey2020}. Polarization measurements enable the discrimination among such models, including cases in which they are degenerate when assessed solely through spectroscopy \cite{Jeffrey2020}. To date, only a few X-ray polarization measurements with low statistical significance have been performed \cite{Tindo1970,Tindo1972a, Tindo1972b,Tramiel1984,SuarezGarcia2006, Boggs2006}. High-significance measurements would overcome such degeneracies\cite{Jeffrey2020}. To advance the understanding of SF physics, CUSP will measure linear polarization in the 25-100 keV energy band.

Furthermore, because of SFs are dynamic events (with timescales ranging from minutes to hours), CUSP will have the capability to study polarization as a function of time. The CUSP science program will also include ancillary science topics, involving the observation of intense X-ray astrophysical sources that may be intercepted by the instrument's Field of View (F.O.V. approximately $\pm36^\circ$) while observing the Sun. These ancillary science topics include serendipitous Gamma-Ray Bursts and other sources such as for example the Crab Nebula (PWN), Sco X-1 (LMXB), and A0535+26 (HMXB).

CUSP is currently approved for a Phase B study by the Italian Space Agency, as part of the Alcor program which aims to develop CubeSat technologies and missions. INAF-IAPS serves as the Prime Contractor and Principal Investigator institution for the scientific payload, with contributions from INAF-OAS Bologna and INAF-OAR. The payload's front-end and back-end electronics will be designed and built by Deda Connect s.r.l.. The 6U XL CubeSat platform will be designed and produced by IMT s.r.l.. Mission analysis is being performed by the Interdepartmental Center for Aerospace Industrial Research (CIRI-AERO) of the University of Bologna, while the University of Viterbo "La Tuscia" will manage the Ground Station and Mission Operation Center (MOC). The Science Operation Center (SOC) could be in charge to the ASI Space Science Data Center (ASI SSDC). The final decision will be finalised in the next phases.

\section{The payload: The hard X-ray polarimeter}
\label{sec:payload}
\begin{figure} [ht]
\begin{center}
\begin{tabular}{c}
\includegraphics[height=10cm]{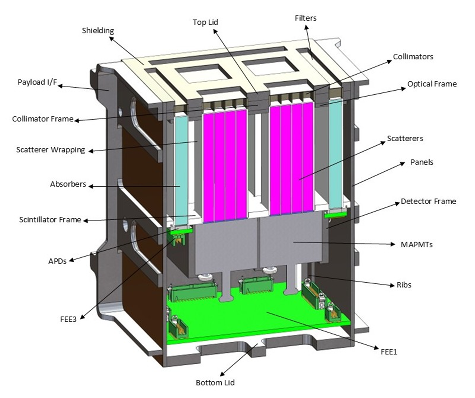}
\end{tabular}
\end{center}
\caption[example]
{Current Phase B design of the CUSP payload.\label{fig:payload}}
\end{figure}

The CUSP payload features a dual-phase Compton scattering polarimeter that operates within the 25-100 keV energy range. To perform polarimetric measurements, the instrument records coincident events between plastic and inorganic scintillator rods (see Fig.~\ref{fig:payload}). The plastic scintillator, because of its low atomic number, maximizes the probability of Compton scattering. In contrast, the heavier inorganic crystal, made of GAGG (Gd$_3$Al$_2$Ga$_3$O$_{12}$), is employed to maximize the photoelectric absorption of the scattered photons. The azimuthal angular distribution of these plastic/GAGG coincidence events is then used to measure the polarization of the incoming radiation. This is possible because polarized radiation causes a preferential azimuthal scattering direction (perpendicular to the incident beam axis), as described by the Klein-Nishina cross-section \cite{Heitler1954}:
\begin{equation}
\frac{d\sigma}{d\Omega}=\frac{{r_0}^2}{2}\frac{{E^\prime}^2}{{E}^2}\Biggr[ \frac{E}{E^\prime}+\frac{E^\prime}{E}-2\sin^2 \theta \cos^2 \phi \Biggl] \label{eq:KN}
\end{equation}
where
\begin{equation}
\frac{E'}{E}=\frac{1}{1+\frac{E}{m_e c^2}(1-\cos \theta)}\label{eq:EsuE}
\end{equation}
Here, $E$ and $E^\prime$ are the respective energies of the incident and scattered photons. The polar scattering angle, $\theta$, is measured from the incident photon's direction, while the azimuthal angle, $\phi$ , is measured from the plane formed by the incoming direction and the incident photon's electric vector.
Linearly polarized photons preferentially scatter perpendicular to their polarization direction. This behavior produces a modulation in the histogram of the azimuthal angle $\phi$. The higher the modulation response to polarised radiation, the higher the polarimeter's sensitivity. Thus, the response to polarised radiaiton is quantified by the modulation factor, $\mu(\theta)$ , which is the fraction of the modulated signal corresponding to 100\% polarized radiation:
\begin{equation}
\mu(\theta)=\frac{N_\mathrm{max}(\theta)-N_\mathrm{min}(\theta)}{N_\mathrm{max}(\theta)+N_\mathrm{min}(\theta)}=\frac{(\frac{d\sigma}{d\Omega})_{\phi=\frac{\pi}{2}}-(\frac{d\sigma}{d\Omega})_{\phi =0}}{(\frac{d\sigma}{d\Omega})_{\phi=\frac{\pi}{2}}+(\frac{d\sigma}{d\Omega})_{\phi =0}}=\frac{\sin^2\theta }{\frac{E}{E^\prime}+\frac{E^\prime}{E}-\sin^2 \theta} \label{eq:Muphi}
\end{equation}
For an energy of the incident photon $E \ll mec^2 = 511$~keV (coherent scattering) $E = E^\prime$ and the maximum modulation factor is obtained for $\theta=90^\circ$ (orthogonally to the incident photons direction). By increasing the energy it occurs at narrower scattering angles (forward folding of scattered photons). However, at 100 keV it is still $\theta \simeq 90^\circ$ and modulation factor $\sim 100\%$\cite{Fabiani2012c}.
The polarimeter sensitivity is given by the Minimum Detectable Polarization (MDP) at 99\% confidence level\cite{Weisskopf2010}:
\begin{equation}
MDP_{99\%}=\frac{4.29}{\mu R}\sqrt{\frac{R+B}{T}} \label{eq:MDP}
\end{equation}
where $R$ is the source count rate, $B$ the background rate and $T$ the observing time.

\begin{figure} [ht]
\begin{center}
\begin{tabular}{c}
\includegraphics[height=8cm]{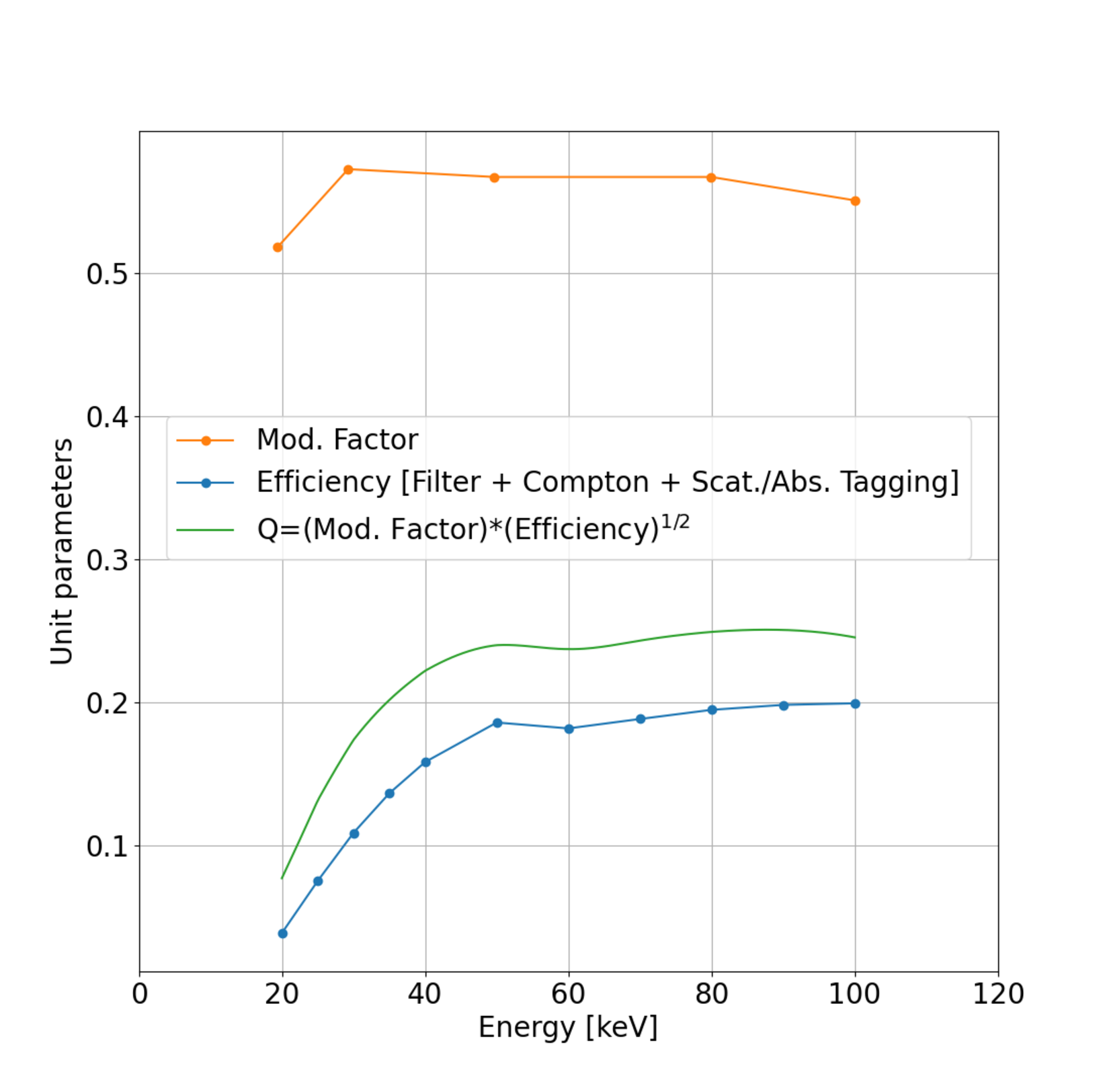}
\end{tabular}
\end{center}
\caption[example]
{Current estimates of the modulation factor, efficiency (Compton interaction and tagging efficiency) and quality factor. \label{fig:curves}}
\end{figure}

The CUSP payload comprises a W collimator system to define the field of view of the scintillator-based detection system comprising plastic scintillator bars read out by four 64-channel Multi-Anode Photomultiplier Tubes (MAPMTs) and GAGG inorganic scintillator bars read out by 32 Avalanche Photodiodes (APDs). 
Figure~\ref{fig:payload} shows the current design of the CUSP Payload. The estimates of the modulation factor $\mu$, quantum efficiency $\epsilon$, and quality factor $Q$, defined as 
\begin{equation}
Q=\mu \sqrt{\epsilon}\label{eq:Q}
\end{equation}
are shown in Figure~\ref{fig:curves}. The quantum efficiency incorporates the tagging efficiency, defined as the probability of detecting an energy deposit in the scatterer, once the scattered photon have been detected in the absorber \cite{Fabiani2012c}. The quality factor determines the optimal energy range for the polarimeter's polarization measurements and is derived from the Minimum Detectable Polarization (MDP) \cite{Weisskopf2010} for a source-dominated observation.

Table \ref{tab:mdp} lists the current best estimate for the Minimum Detectable Polarization (MDP) in the 25-100~keV energy band, derived from benchmark solar flares in Saint-Hilaire et al. (2008) \cite{SaintHilaire2008}. Since the polarization of solar flares is expected to be significant (on the order of tens of percent), it is well above this MDP, allowing for a confident measurement with only a few minutes of integration time.
\begin{table}[ht]
\caption{Current best estimate (end of Phase A) of the MDP in the 25-100~keV energy band based on benchmark solar flares from Saint-Hilaire et al. (2008)\cite{SaintHilaire2008}}
\label{tab:mdp}
\begin{center}      
\begin{tabular}{|c|c|c|}
\hline
\rule[-1ex]{0pt}{3.5ex} Flare Class & Integration Time (s) & MDP ($\%$)  \\
\hline
\hline
\rule[-1ex]{0pt}{3.5ex} M 5.2 & 284 & 7.8  \\
\hline
\rule[-1ex]{0pt}{3.5ex} X 1.2 & 240 & 3.9  \\
\hline
\rule[-1ex]{0pt}{3.5ex} X 10 & 351 & 0.9  \\
\hline
\end{tabular}
\end{center}
\end{table}

\section{THE MISSION CONCEPT}
The CUSP mission underwent a trade-off analysis during its Phase B, which was consolidated by the conclusion of the System Requirements Review in July 2025. Following this analysis, the Italian Space Agency selected as a baseline a single-satellite mission configuration over a two-satellite constellation. The proposed constellation would have consisted of two satellites in the same Sun-Synchronous Orbit (SSO) at an altitude of approximately 500 km.

A detailed mission analysis was performed for a 6U XL CubeSat platform without a propulsion system to determine the total observation time of the Sun during the three years of niminal mission duration. It was determined that if two satellites were launched with an initial phase difference of $180^\circ$, this separation would be quickly lost within a few months. The primary causes of the phase lost are the orbital perturbations, dominated by atmospheric drag, and the uncertainties in the orbital elements upon deployment.
To evaluate the different mission configurations the mission analysis assessed orbits with three different Local Times of the Ascending Node (LTAN) to calculate the percentage of solar observation time and, consequently, the number of detectable solar flares:
\begin{itemize}
    \item in a Noon-Midnight (12:00 LTAN) or a Mid-Morning (10:00 LTAN) orbit, a single satellite observes the Sun for approximately $45\%$ of the time. A two-satellite constellation would increase this to about $68\%$.
    \item in a Dawn-Dusk (06:00 LTAN) orbit, a single satellite can observe the Sun for roughly $69\%$ of the time, while a two-satellite constellation would increase this to approximately $88\%$.
\end{itemize}
These observation times translate to an increase in the number of observable solar flares of about $50\%$ for the Noon-Midnight and Mid-Morning orbits, and approximately $30\%$ for the Dawn-Dusk orbit.
Furthermore, in a two-satellite constellation, the Sun would have been simultaneously observable for about $22\%$ of the time in the Noon-Midnight/Mid-Morning orbits and $51\%$ in the Dawn-Dusk orbit. This simultaneous observation would have improved the Minimum Detectable Polarization (MDP) by a factor of $\sqrt{2}\sim 40\%$ and would have allowed for a direct comparison of polarization results from a solar flare observed by both satellites.

The trade-off analysis concluded that while a two-satellite constellation offers a significant increase in the number of observable solar flares, it does not enable any science objectives that cannot be fundamentally addressed by a single satellite. The only scenario with a significant advantage is the improved MDP, which would enable the measurement of the low degree (a few percent) of polarization expected from thermal emission models of solar flares, at least for X-class events \cite{Emslie1980a}. However, a single satellite remains capable of measuring with high significance the higher polarization degrees (tens of percent) expected from non-thermal models, thereby distinguishing them from thermal emissions, for which only upper limits of the polarization degree could be established.

\begin{figure} [ht]
\begin{center}
\begin{tabular}{c}
\includegraphics[height=8cm]{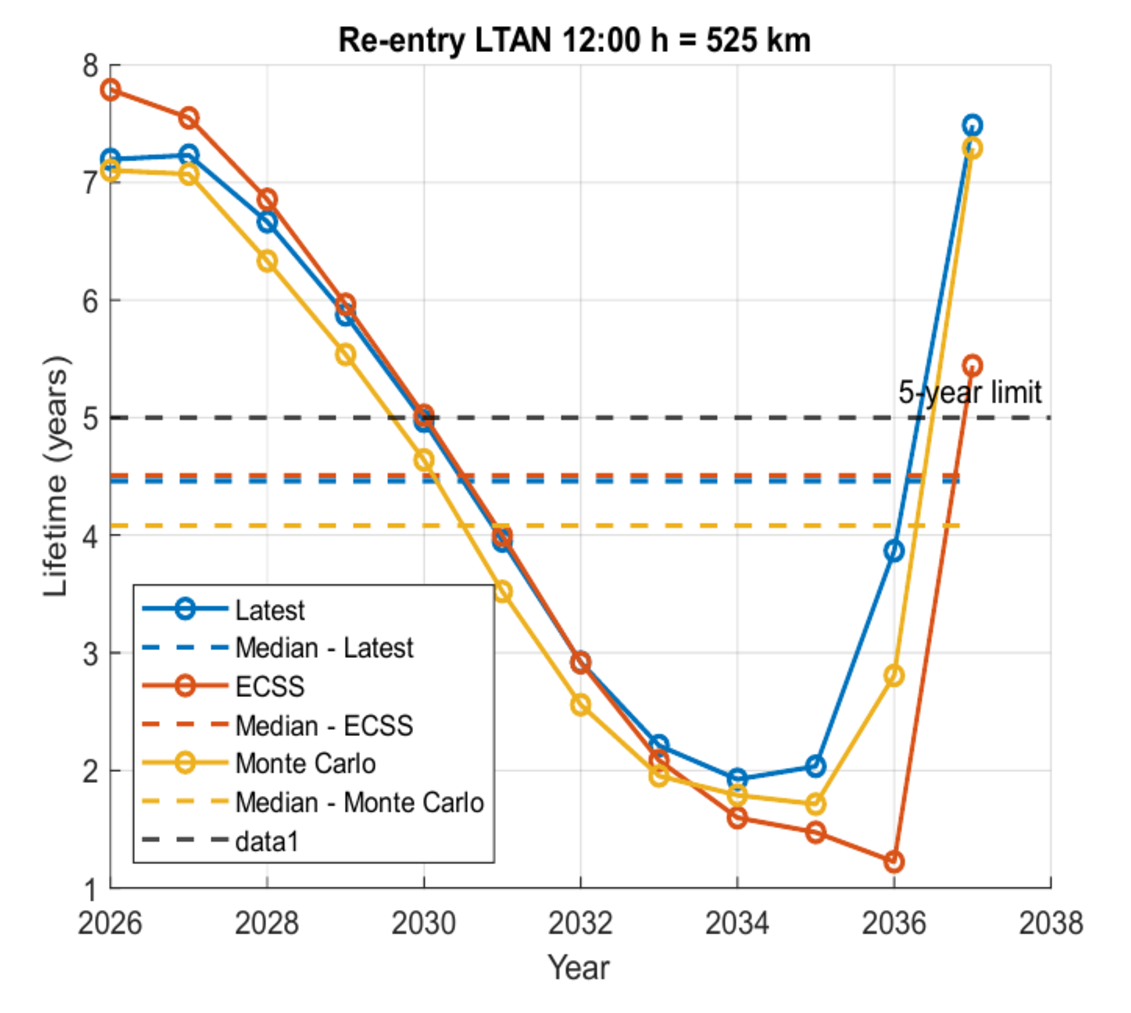}
\end{tabular}
\end{center}
\caption[example]
{Example of orbit lifetime evolution as a function of launch date and initial height of 525 km for LTAN 12:00 (Noon-Midnight orbit) computed as per ESA debris mitigation guidelines. The 50th percentile of reentry times from different solar activity models are reported\label{fig:reentry}}
\end{figure}

The mission analysis carried out included also a detailed reentry analysis of the satellite based on the most recent ESA guidelines which clarified the interpretation and verification approach for the requirements to be fulfilled by any non-maneuvering s/c orbiting in the LEO protected region\footnote{ESSB-ST-U-007 Issue 1 - ESA Space Debris Mitigation Requirements - ESA Space Debris Mitigation
WG – 30/10/2023 and ESA-TECQI-HO-2024-002968 - Presentation on ESA’s Space Debris Mitigation Requirements -
07/10/2024}.
Two clearance requirements are applicable to CUSP:
\begin{itemize}
    \item the orbit lifetime is less than 5 years starting from the orbit injection epoch;
    \item the cumulative collision probability from its end of life until re-entry with space objects larger
than 1 cm is below $10^{-3}$
\end{itemize}
The orbit lifetime was assessed probabilistically, including at least the variability by moving the starting point through a full solar cycle and selecting the launch epoch with yearly steps. The 50th-percentiles of reentry time were compared to the 5-year limit in case of different solar activity models to increase confidence in the results. In Fig.\ref{fig:reentry} the case of an SSO orbit 525 km of height for LTAN 12:00 (Noon-Midnight orbit) is shown as an example. Due to the fact that the comparison with respect to the 5-year limit is done based on the 50th-percentiles of reentry time, the predicted natural decay for a selected launch epoch around the sola rminimum is larger than 5 years. The current timeline of the mission foresees a launch late 2027/early 2028, by assuming final approval of the mission by ASI with a key decision expected after the closure of Phase B.

During solar observations, the satellite rotates around its sun-pointing axis. This rotation is crucial for reducing systematic errors known as spurious modulation.
The peak intensity of a hard X-ray solar flare typically lasts for only a few minutes. Therefore, a rotation speed of at least 1~RPM is used. For data downlink, the observations are temporarily interrupted as the satellite switches to a 3-axis stabilized configuration to optimize the transmission link.

\section{THE PLATFORM}

\label{sec:platform}  
\begin{figure} [ht]
\begin{center}
\begin{tabular}{c}
\includegraphics[height=7cm]{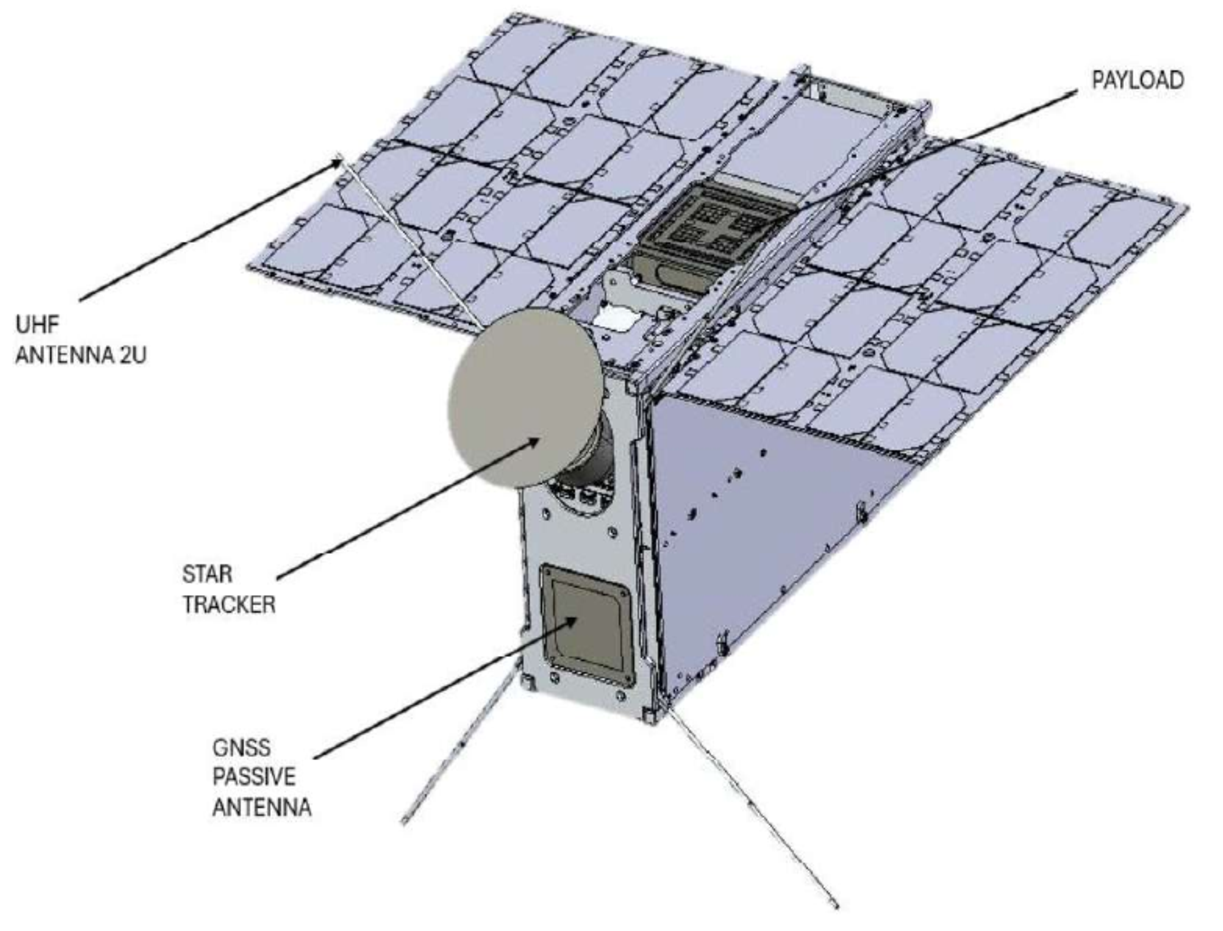}
\end{tabular}
\end{center}
\caption[example]
{Representation of the CUSP satellite. \label{fig:sat}}
\end{figure}
The 6U XL CubeSat platform architecture (see Firg.\ref{fig:sat}) was developed by IMT s.r.l. thanks to the company's extensive experience in the CubeSat field.
The preliminary performance specifications of the CUSP platform are summarized in Table~\ref{tab:platform}. It features a high-performance On-Board Computer (OBC) with multiple communication interfaces for the payload. The OBC includes on-board memory for storing payload data. Furthermore, an additional memory unit is integrated within the Communication module to buffer data prior to its transmission to the Ground Station.
For scientific data acquisition, the satellite enters a sun-pointing mode, rotating around the main payload axis. In this mode, attitude determination must be maintained with an accuracy better than $\pm 2^\circ$. This precision is achieved using a high-precision gyroscope integrated into the Attitude Determination and Control System (ADCS). During non-observational phases, the platform reverts to a three-axis stabilized mode, during which it can downlink collected data to the ground station. Telemetry and Telecommand (TT\&C) operations are handled via a UHF-band link using omnidirectional turnstile antennas. To facilitate the transfer of large volumes of scientific data, an S-band communication subsystem complements the UHF link, offering a significantly higher data rate. The platform operates in various modes depending on the mission state/phase.

Max throughput up to 10 Mbps, nominal ominal up to 5 Mbps.

\begin{table}[ht]
\caption{Preliminary performance of the CUSP platform.}
\label{tab:platform}
\begin{center}      
\begin{tabular}{|c|c|}
\hline
\rule[-1ex]{0pt}{4.5ex} Parameter & Value  \\
\hline
\hline
\rule[-1ex]{0pt}{4.5ex} Peak power & $\sim30$W with deployable panels in Sun Pointing \\
\hline
\rule[-1ex]{0pt}{4.5ex} Battery & 84 Wh  \\
\hline
\rule[-1ex]{0pt}{4.5ex} Attitude accuracy & \parbox{5cm}{$<0.04$ deg @ $3\sigma$ (AKE) \\ $<0.08$ deg @ $3\sigma (APE)$  \\ $<2$ deg/s Slew Rate} \\
\hline
\rule[-1ex]{0pt}{4.5ex} Operative frequencies &  \parbox{5cm}{S-Band downlink \\ UHF-Band uplink / downlink}\\
\hline
\rule[-1ex]{0pt}{4.5ex} Downlink throughputs &   \parbox{5cm}{Up to 10 Mbps (max) \\ Up to 5 Mbps (nominal)}  \\
\hline
\rule[-1ex]{0pt}{4.5ex} Available interfaces & CAN Bus, I2C, UART, SPI, RS485 \\
\hline
\rule[-1ex]{0pt}{4.5ex} Regulated Bus &  3.3V / 5V / 12V  \\
\hline
\rule[-1ex]{0pt}{4.5ex} Unregulated Bus &  16V (12V-16.2V) \\
\hline
\rule[-1ex]{0pt}{4.5ex} Nominal life time &  3 years (LEO) \\
\hline
\end{tabular}
\end{center}
\end{table}

\section{The Ground Segment}
\label{sec:gs}  
\begin{figure} [ht]
\begin{center}
\begin{tabular}{c}
\includegraphics[height=3cm]{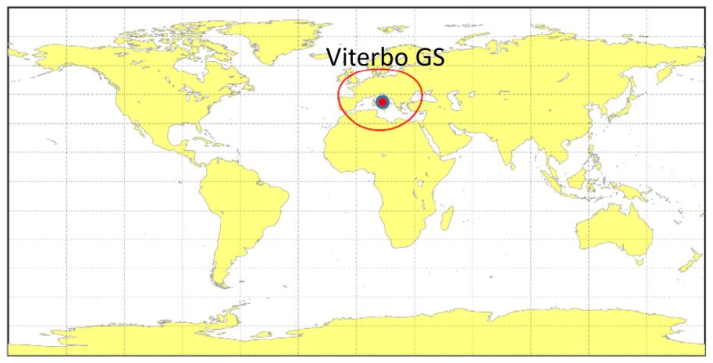}
\includegraphics[height=6cm]{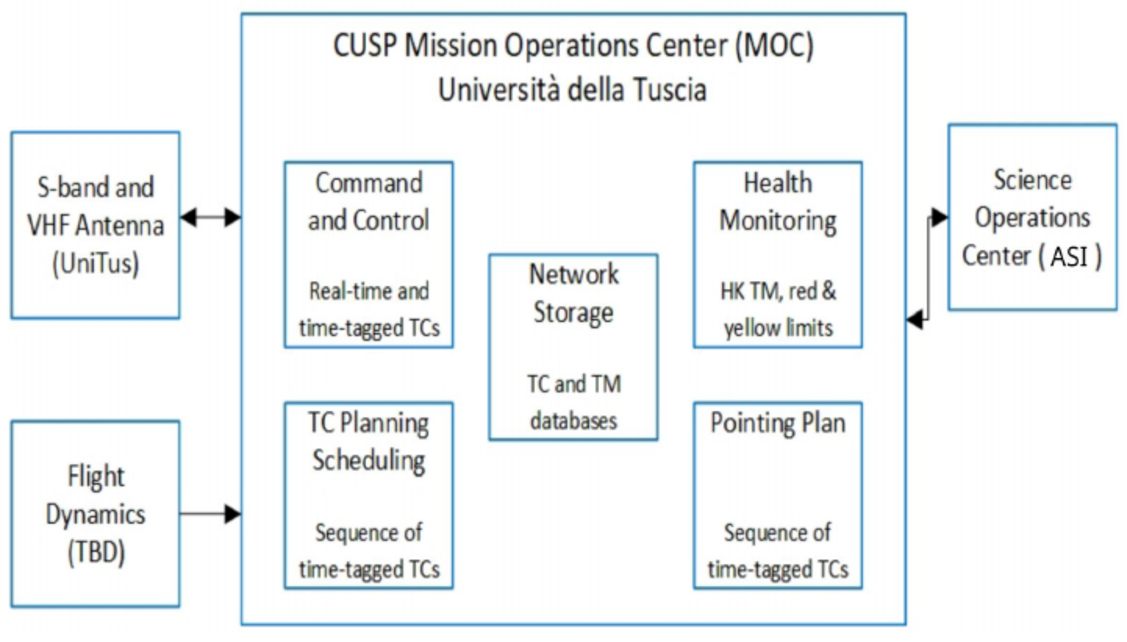}
\end{tabular}
\end{center}
\caption{In the left panel: the ground station at the "La Tuscia" University of Viterbo. In the right panel: the organization of MOC/SOC. \label{fig:groundstation}}
\end{figure}

The CUSP Ground Segment consists of three main components: the Primary Ground Station and the Mission Operations Center (MOC), both located in Building F of the "Riello" Campus at the University of "La Tuscia", Viterbo; and the Science Operations Center (SOC), possibly located at SSDC at the ASI headquarters in Rome. The involvement of ASI/SSDC in the CUSP mission as SOC will be finalised in the next phases.
The Primary Ground Station (see Fig.~\ref{fig:groundstation}) located at the University of "La Tuscia" was built in 2019 as part of the HORTA project (Italian regional funds POR-FESR 2014-2020 of Lazio region). The Ground Station allows for autonomous satellite tracking (using TLE satellite data - Two Lines Elements) and satellite communication.
The available antennas and bands are:
\begin{itemize}
\item VHF: Uplink and Downlink
\item UHF: Uplink and Downlink
\item S-band: Downlink
\end{itemize}
The UHF/VHF bandwidth are 9.6 kbps as default for downlink (available also 1.2/ 2.4 / 4.8 kbps)
and 1.2 kbps as default for uplink (available also 2.4 / 4.8/ 9.6 kbps).
The S-band bandwidth is up to 1 Mbps for downlink.
The pointing accuracy of the ground station is 0.1° (both azimuth and elevation) with a minimum tracking speed of 2$^\circ$/sec in azimuth, 1.8$^\circ$/sec in elevation.
Telemetry and Telecommand (TT\&C) for the CUSP mission satellites will be handled via UHF band uplink and downlink communications using omnidirectional turnstile antennas.
To allow scientific data transfer, the S-band communication subsystem supplements the UHF link, providing a higher downlink speed.
Data received from the ground station is transferred via fibre optics cable to dedicated workstations in the MOC that allows to schedule the passage of the satellite, the
TT\&C and Payload data transceiving activities. Moreover, it provides a Network Server service for data delivery to third parties. The MOC and SOC are linked via an internet connection. 
Backup ground stations, which will be selected in later stages of the project, will also connect to the MOC via the internet. 
The average contact time with the Primary Ground Station is about 5.5 minutes for the range of SSO orbits of interest of the CUSP mission at about 500 km of height. In the worst case scenario of 2 contacts
per day (average contact per day about 3.5) allows to downlink an X10 solar flare within 48h.

All data delivered from the Mission Operations Center (MOC) will be processed and archived at the SOC. Based on a decision to be finalised in the next phases, CUSP mission could take advantage of ASI Space Science Data Center (ASI/SSDC) as SOC. It is a research infrastructure established in 2016 at ASI Headquarters. It was established  from the heritage of scientific data management that dates back to 2000. The center has since broadened its scope from Astrophysics to Earth Observation and information technologies for large-scale data management. Its core mission is to provide services that maximize data usability and ensure interoperability with other data centers worldwide, adhering to internationally recognized standards.
In case of ASI/SSDC will be in charge of the SOC, then INAF, which also contributes to its the ordinary activities, will collaborate on developing the scientific data processing pipeline. This collaboration will leverage INAF's instrumental knowledge and its experience in managing X-ray polarimetric data, gained from past space missions such as IXPE\cite{Weisskopf2025}.

\section{The planning}
\label{sec:planning}

The CUSP project entered its 12-month Phase B in December 2024. In addition to ordinary phase B activities, the CUSP phase B also includes the integration and testing of a Structural Model (SM) and a Front-End Detector Prototype (FEDP). The SM replicates the mechanical structure of the flight polarimeter holding the scintillator bars; it will be used to verify the design's stiffness and its ability to withstand launch loads by keeping the bars in place.
The FEDP is designed to demonstrate the capability to perform coincidence measurements using both SKIROC and MAROC ASICs, which are managed by an FPGA. The prototype's front-end electronics, including the interface electronics for the sensors (MAPMT and APDs), are representative of the flight model. These activities are expected to raise the detector's Technology Readiness Level (TRL) from 3 to 4.

Parallel to these prototyping efforts, an Engineering Qualification Model (EQM) of the payload is being designed during Phase B. The EQM will be integrated and tested during the subsequent Phase C to advance the detector TRL from 4 to 7. Regarding the satellite platform, a single CubeSat Proto-Flight Model (PFM) will be built and qualified.

The Calibration of the Hard X-ray Polarimeter of each CubeSat will be carried out at INAF-IAPS calibration facility (already employed for calibrating the IXPE Detector Units)\cite{Muleri2021ice} and possibly to Synchrotron test beamlines. The INAF-IAPS calibration facility is going to be extended up to hard X-rays by employing X-ray tubes up to 100-150 KV for direct beams and scattered ones (polarized radiation)

\acknowledgments 
 
This work is funded by the Italian Space Agency (ASI) within the Alcor Program, as part of the development of the CUbesat Solar Polarimeter (CUSP) mission under ASI-INAF contract n. 2023-2-R.0.  


\begin{thebibliography}{10}

\bibitem{Papaioannou2016}
{Papaioannou}, A., {Sandberg}, I., {Anastasiadis}, A., {Kouloumvakos}, A., {Georgoulis}, M.~K., {Tziotziou}, K., {Tsiropoula}, G., {Jiggens}, P., and {Hilgers}, A., ``{Solar flares, coronal mass ejections and solar energetic particle event characteristics},'' {\em Journal of Space Weather and Space Climate}~{\bf 6},  A42 (Dec. 2016).

\bibitem{Emslie1980a}
{Emslie}, A.~G. and {Brown}, J.~C., ``{The polarization and directivity of solar-flare hard X-ray bremsstrahlung from a thermal source},'' {\em \apj}~{\bf 237},  1015--1023 (May 1980).

\bibitem{Nagasawa2022}
{Nagasawa}, S., {Kawate}, T., {Narukage}, N., {Takahashi}, T., {Caspi}, A., and {Woods}, T.~N., ``{Study of Time Evolution of Thermal and Nonthermal Emission from an M-class Solar Flare},'' {\em \apj}~{\bf 933},  173 (July 2022).

\bibitem{Zharkova2010}
{Zharkova}, V.~V., {Kuznetsov}, A.~A., and {Siversky}, T.~V., ``{Diagnostics of energetic electrons with anisotropic distributions in solar flares. I. Hard X-rays bremsstrahlung emission},'' {\em \aap}~{\bf 512},  A8 (Mar. 2010).

\bibitem{Grigis2004}
{Grigis}, P.~C. and {Benz}, A.~O., ``{The spectral evolution of impulsive solar X-ray flares},'' {\em \aap}~{\bf 426},  1093--1101 (Nov. 2004).

\bibitem{Dennis2005}
{Dennis}, B.~R., {Phillips}, K.~J.~H., {Sylwester}, J., {Sylwester}, B., {Schwartz}, R.~A., and {Tolbert}, A.~K., ``{The thermal X-ray spectrum of the 2003 April 26 solar flare},'' {\em Advances in Space Research}~{\bf 35},  1723--1727 (Jan. 2005).

\bibitem{Jeffrey2020}
{Jeffrey}, N. L.~S., {Saint-Hilaire}, P., and {Kontar}, E.~P., ``{Probing solar flare accelerated electron distributions with prospective X-ray polarimetry missions},'' {\em \aap}~{\bf 642},  A79 (Oct. 2020).

\bibitem{Tindo1970}
{Tindo}, I.~P., {Ivanov}, V.~D., {Mandel'Stam}, S.~L., and {Shuryghin}, A.~I., ``{On the Polarization of the Emission of X-Ray Solar Flares},'' {\em \solphys}~{\bf 14},  204--207 (Sept. 1970).

\bibitem{Tindo1972a}
{Tindo}, I.~P., {Ivanov}, V.~D., {Mandel'Stam}, S.~L., and {Shuryghin}, A.~I., ``{New Measurements of the Polarization of X-Ray Solar Flares},'' {\em \solphys}~{\bf 24},  429--433 (June 1972).

\bibitem{Tindo1972b}
{Tindo}, I.~P., {Ivanov}, V.~D., {Valn{\'{\i}}{\v c}ek}, B., and {Livshits}, M.~A., ``{Preliminary Interpretation of the Polarization Measurements Performed on 'Intercosmos-4' during Three X-Ray Solar Flares},'' {\em \solphys}~{\bf 27},  426--435 (Dec. 1972).

\bibitem{Tramiel1984}
{Tramiel}, L.~J., {Novick}, R., and {Chanan}, G.~A., ``{Polarization evidence for the isotropy of electrons responsible for the production of 5-20 keV X-rays in solar flares},'' {\em \apj}~{\bf 280},  440--447 (May 1984).

\bibitem{SuarezGarcia2006}
{Suarez-Garcia}, E., {Hajdas}, W., {Wigger}, C., {Arzner}, K., {G{\"u}del}, M., {Zehnder}, A., and {Grigis}, P., ``{X-Ray Polarization of Solar Flares Measured with Rhessi},'' {\em \solphys}~{\bf 239},  149--172 (Dec. 2006).

\bibitem{Boggs2006}
{Boggs}, S.~E., {Coburn}, W., and {Kalemci}, E., ``{Gamma-Ray Polarimetry of Two X-Class Solar Flares},'' {\em \apj}~{\bf 638},  1129--1139 (Feb. 2006).

\bibitem{Heitler1954}
{Heitler}, W.,  [{\em {Quantum theory of radiation}}{\nolinebreak\hspace{0.1em}]}, International Series of Monographs on Physics, Oxford: Clarendon, 1954, 3rd ed. (1954).

\bibitem{Fabiani2012c}
{Fabiani}, S., {Campana}, R., {Costa}, E., {Del Monte}, E., {Muleri}, F., {Rubini}, A., and {Soffitta}, P., ``{Characterization of scatterers for an active focal plane Compton polarimeter},'' {\em Astroparticle Physics}~{\bf 44},  91--101 (Apr. 2013).

\bibitem{Weisskopf2010}
{Weisskopf}, M., {Elsner}, R., and {O'Dell}, S., ``{On understanding the figures of merit for detection and measurementof x-ray polarization},'' {\em Proceedings of the SPIE,}~{\bf 7732},  77320E--77320E--5 (2010).

\bibitem{SaintHilaire2008}
{Saint-Hilaire}, P., {Krucker}, S., and {Lin}, R.~P., ``{A Statistical Survey of Hard X-ray Spectral Characteristics of Solar Flares with Two Footpoints},'' {\em \solphys}~{\bf 250},  53--73 (July 2008).

\bibitem{Weisskopf2025}
{Weisskopf}, M. and {Soffitta}, P., ``{The Imaging X-ray Polarimetry Explorer (IXPE)},'' in [{\em American Astronomical Society Meeting Abstracts \#245}{\nolinebreak\hspace{0.1em}]},  {\em American Astronomical Society Meeting Abstracts} {\bf 245},  300.01 (Jan. 2025).

\bibitem{Muleri2021ice}
{Muleri}, L. and {et al.}, ``{The IXPE Instrument Calibration Equipment},'' {\em In Preparation}~{\bf 00},  00--00 (Dec. 2021).

\end{thebibliography}
\bibliographystyle{spiebib} 

\end{document}